\documentclass[twocolumn,preprintnumbers,amsmath,amssymb,superscriptaddress,longbibliography]{revtex4-2}
\usepackage{graphicx}
\usepackage{multirow}
\usepackage{amsmath}
\usepackage{dcolumn}
\usepackage{bm}
\usepackage[usenames,dvipsnames]{xcolor}
\usepackage{comment}

\begin{document}
\preprint{\today}

\title{Droplets sliding on single and multiple vertical fibers}

\author{M. Leonard}
  \affiliation{GRASP, Physics Department, University de Li\`ege, Belgium.}
\author{J. Van Hulle}
\affiliation{GRASP, Physics Department, University de Li\`ege, Belgium.}
\author{F. Weyer}
\affiliation{GRASP, Physics Department, University de Li\`ege, Belgium.}
\author{D. Terwagne}
\affiliation{Frugal LAB, Physics Department, Université Libre de Bruxelles, Belgium.}
\author{N. Vandewalle}
\affiliation{GRASP, Physics Department, University de Li\`ege, Belgium.}


\begin{abstract} 
From microfluidics to fog harvesting applications, tiny droplets are transportedalong various solid substrates including hairs, threads, grooves and other light structures. Driven by gravity, a droplet sliding along a vertical fiber is a complex problem since it is losing volume and speed as it goes down. With the help of an original setup, we solve this problem by tracking in real time droplet characteristics and dynamics. Single fibers as well as multiple fiber systems are studied to consider the presence of grooves. On both fibers and grooved threads, droplet speed and volume are seen to decay rapidly because the liquid entity is leaving a thin film behind. This film exerts a capillary force able to stop the droplet motion before it is completely drained. A model is proposed to capture the droplet dynamics. We evidence also that multiple vertical fibers are enhancing the droplet speed while simultaneously promoting increased liquid loss on grooves. 
\end{abstract}

\maketitle

\section{Introduction}

In arid or semi-arid regions, evolution of diverse flora and fauna has developed various survival strategies to access water despite the challenging conditions. Many of these living organisms utilize naturally occurring atmospheric water, either through fog collection \cite{ebner2011efficient, andrews2011three, pan2016upside, ito2015mechanics} or promoting condensation \cite{ju2012multi}. Once collected, water must be transported to the area where the organism can absorb it. To achieve this, nature primarily relies on three fundamental principles: gravity \cite{roth2012leaf}, specific geometries, and wettability gradients \cite{guadarrama2014dew}, often in combination. In particular, specific geometries like conical shapes \cite{ju2012multi,chen2018ultrafast,van2021capillary} or surfaces with specifically arranged substructures such as grooves \cite{chen2018ultrafast} and crevaces \cite{shi2022bioinspired} generate capillary forces.

If we consider gravity-driven transport, the dynamics of drops on vertical or inclined fibers is a complex phenomenon \cite{huang2009equilibrium}.
Indeed, depending on the inclination fiber angle \cite{gilet2010droplets}, the fiber diameter \cite{mchale2001shape} and surface tension \cite{gabbard2021asymmetric}, the droplet may adopt two different shapes, barrel or clam, having different contact lines and therefore different dissipation mechanisms. The crossing of fibers has also been extensively studied both from a static \cite{duprat2012wetting, sauret2015wetting} and dynamic \cite{gilet2009digital} view point. Notably, it has been shown in the dynamic case that a fiber crossing can stop or divide the droplet into small volumes \cite{gilet2009digital}. The geometry of the crossing is the major parameter for determining the maximum volume remaining there \cite{weyer2015compound,Pan2018}. A fiber array with selected fiber diameters is able to control the path of droplets \cite{weyer2017switching}. When placing different immiscible liquids on a fiber array, multiple component droplets can be created, opening ways to microfluidic reactors and devices for probing biochemical reactions inside droplets. Several works have also proposed water harvesting using nets \cite{park2013optimal} or harps \cite{shi2018fog,shi2020harps,Jiang2023} which are constituted by vertical fibers transporting the droplets to a reservoir thanks to gravity. The dynamics of droplets sliding on fibers is therefore a key element for many applications from microfluidics to fog harvesting devices. 

To enhance transport efficiency, nature frequently arranges itself into multiple levels of structures. For instance, on the spines of a cactus, there are grooves, hairs, and spikes \cite{ju2012multi}, while spider webs feature silk beads \cite{zheng2010directional}. These characteristics passively facilitate liquid transport through capillary driving forces \cite{Kolliopoulos2021, VanHulle2023}.

\begin{figure}[h]
\begin{center}
\includegraphics[width=0.45\textwidth]{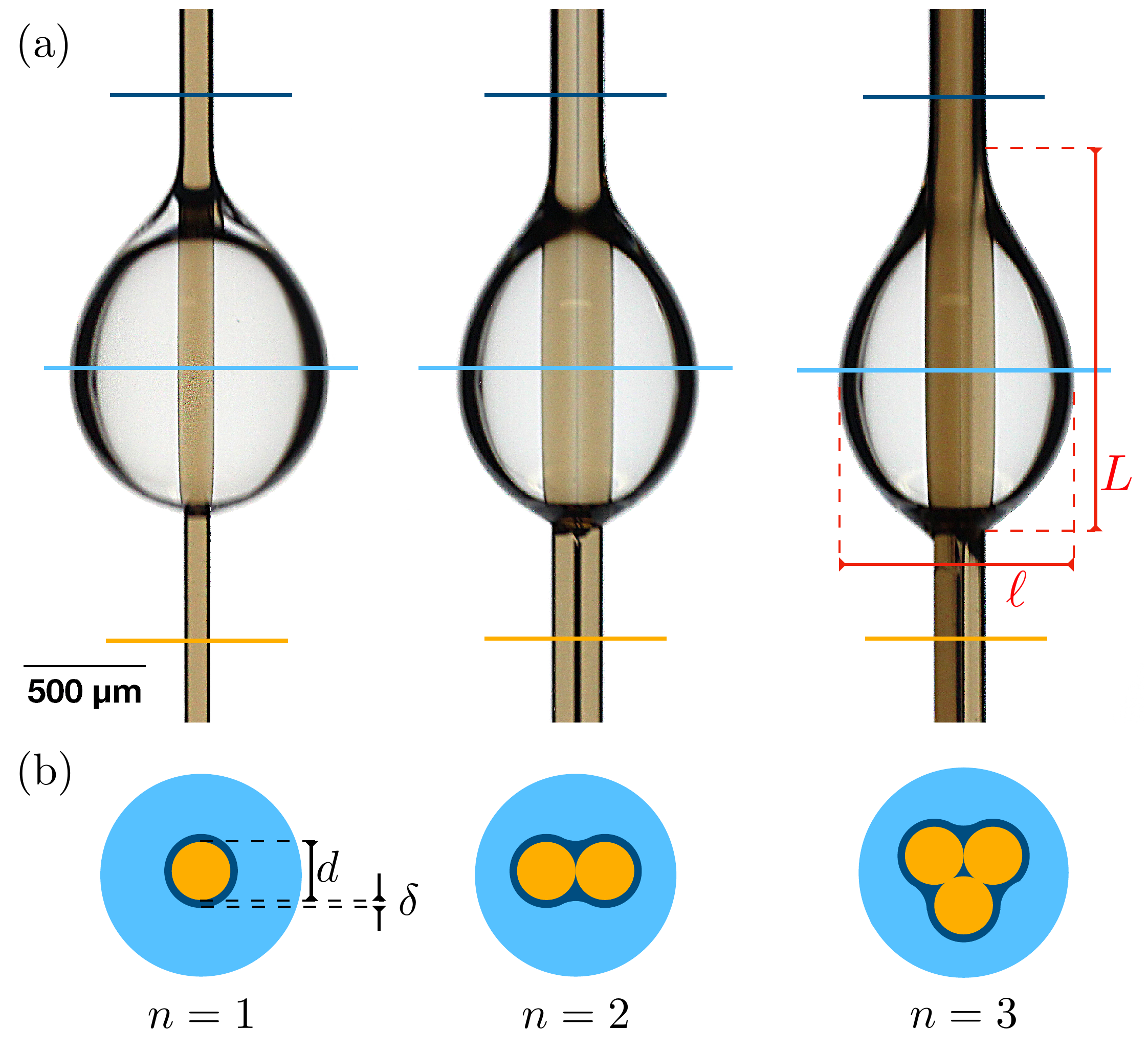}
\caption{(a) Experimental pictures of droplets of $\rm{3 \mu l}$ sliding down 1, 2 and 3 fibers of $d= \rm{140 \mu m}$ from left to right. A liquid film is seen behind the droplet. Droplet shape characteristics : height $L$ and width $\ell$ are emphasized in red. (b) Sketch of three horizontal cuts of the system : dry fibers in front of the droplet (orange), the droplet cross section (light blue) and the liquid film of thickness $\delta$ after the passage of the droplet (dark blue).}
\label{Figure_droplets}
\end{center}
\end{figure}

To add some texture to the vertical fibers, we consider a bundle of fibers with grooves by assembling several fibers together, as shown in Figure \ref{Figure_droplets}(a). This figure shows pictures of droplets sliding on $n=1$, 2 and 3 adjacent vertical fibers of $\rm{140 \mu m}$ in diameter. One observes that the shape of each droplet is significantly different at the bottom and top : while the fiber is dry before the passage of the droplet, a liquid film is left behind. The coating of the fiber is well seen in Figure \ref{Figure_droplets}(a) and is sketched in Figure \ref{Figure_droplets}(b). The presence of this liquid film explains partially the difference of local curvatures at the front and at the rear of the droplet. Because of the contact between the different fibers, grooves are formed and more liquid is expected to accumulate within. This may also change the droplet shape as seen depend on the number $n$ of fiber in Figure \ref{Figure_droplets}

A fundamental question, that we address in the present paper, is to estimate the impact of these substructures on the droplet speed and liquid coating. This study represents a challenge since multiple physical characteristics of the droplet such as speed and volume should be measured synchronously. Thanks to an original setup detailed below, we overcome this experimental difficulty, allowing for the modeling of moving droplets on fibers.

\section{Material and methods}

Nylon fibers are used with diameters ranging from $d=\rm{80 \mu m}$ to $d=\rm{280 \mu m}$. Bundles from $n=1$ to $n=4$ fibers were fabricated. Larger $n$ values have been tested but the entire thread becomes so large that droplets lose their axisymmetric shape as described by Gabbard and Bostwick \cite{gabbard2021asymmetric,gabbard2023bead}. To guarantee fiber-to-fiber contact within the bundles, a slight torsion is applied, with a wavelength significantly larger than the droplet size. This ensures that the torsion does not interfere with the behavior of the droplets. By varying the number $n \le 4$ of fibers, we modify the substructure of the vertical threads, forming $n_g$ grooves. Please note that different configurations may form when $n$ increases. In particular, for $n=4$ two configurations can be formed but they possess similar characteristics. The effective diameter $d_e$ of the structure is considered by comparing the external perimeter of the grooved structure with an equivalent cylinder. The table \ref{tab:bundles} is summarizing the equivalent diameter $d_e$ and groove number $n_g$ for the structures studied herein. 

\begin{table}[h]
\caption{\label{tab:bundles}Main characteristics of fiber bundles considered herein : effective diameter $d_e$ and groove number $n_g$ as a function of the number of fibers $n$.  }
\begin{ruledtabular}
\begin{tabular}{llll}
sketch & $n$ & $d_e$ & $n_g$ \\ \hline
\includegraphics[width=0.04\textwidth]{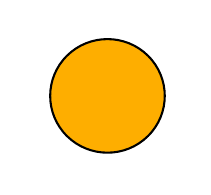} & 1 &  $d$ & 0 \\
\includegraphics[width=0.05\textwidth]{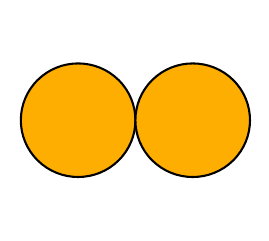} & 2  &  $2d$  & 2 \\
\includegraphics[width=0.05\textwidth]{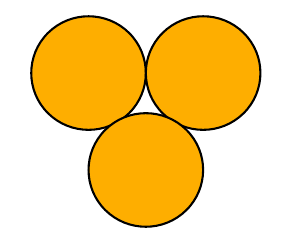} & 3 &  $5d/2$ & 3 \\ 
\includegraphics[width=0.1\textwidth]{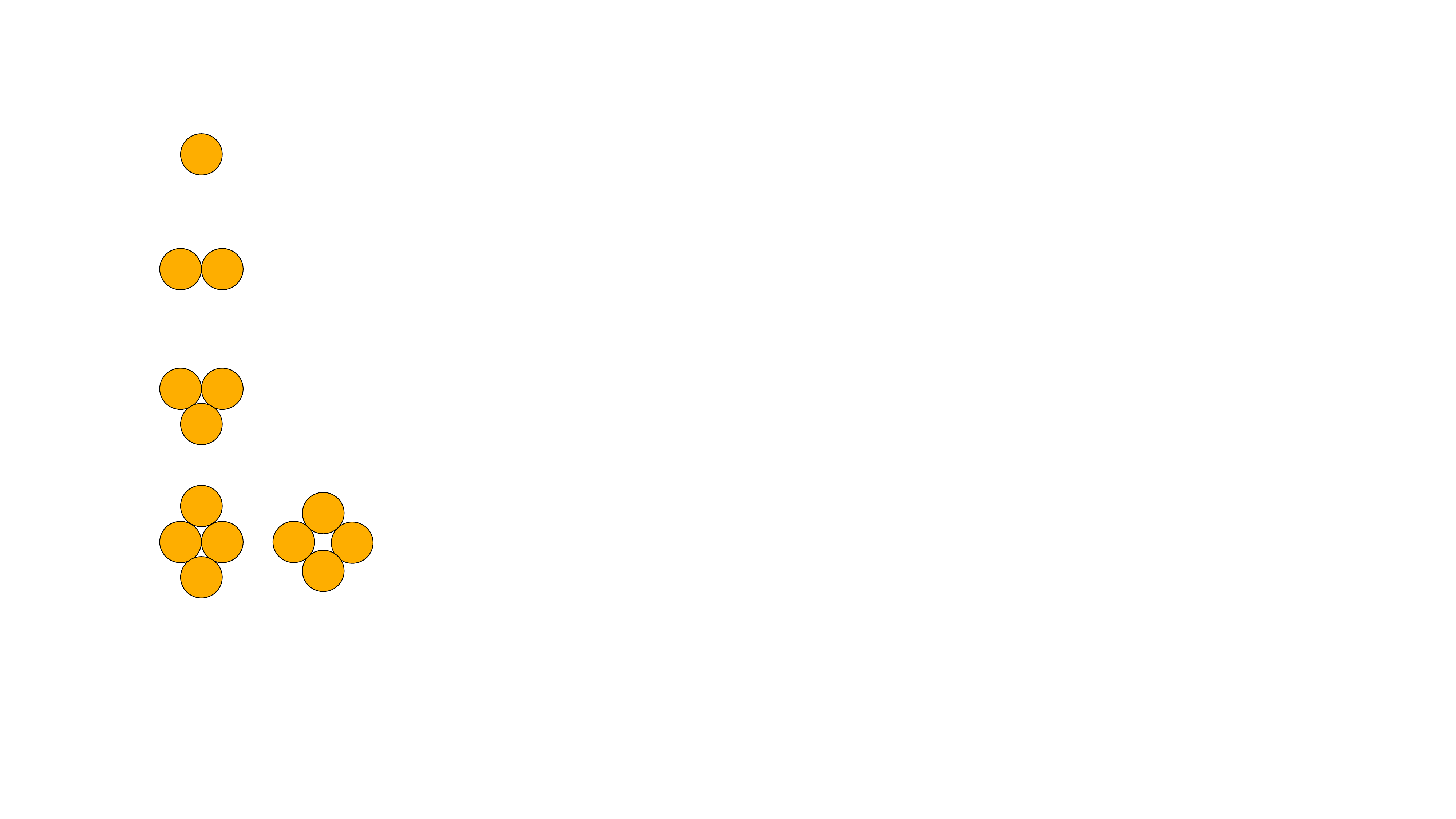} & 4 & $3d$ & 4 \\
\end{tabular}
\end{ruledtabular}
\end{table} 

On these fibers/bundles, we placed $\rm{3 \mu l}$ volume droplets, thanks to a micropipette. We chose silicone oils to ensure the total wetting of the fibers. However, changing $n$ will modify the wetting conditions as observed in Figure \ref{Figure_droplets}. The oil viscosity is in the range $\eta = 10^{-2}$ to $ 5\times 10^{-2}\, \rm{Pa \, s}$, its density is $\rho = \rm{940 \, kg/m^3}$ and its surface tension at $25^{\circ}$C is $20\times 10^{-3}\, \rm{N /m}$.

\begin{figure}[h]
\begin{center}
(a) \includegraphics[width=0.40\textwidth]{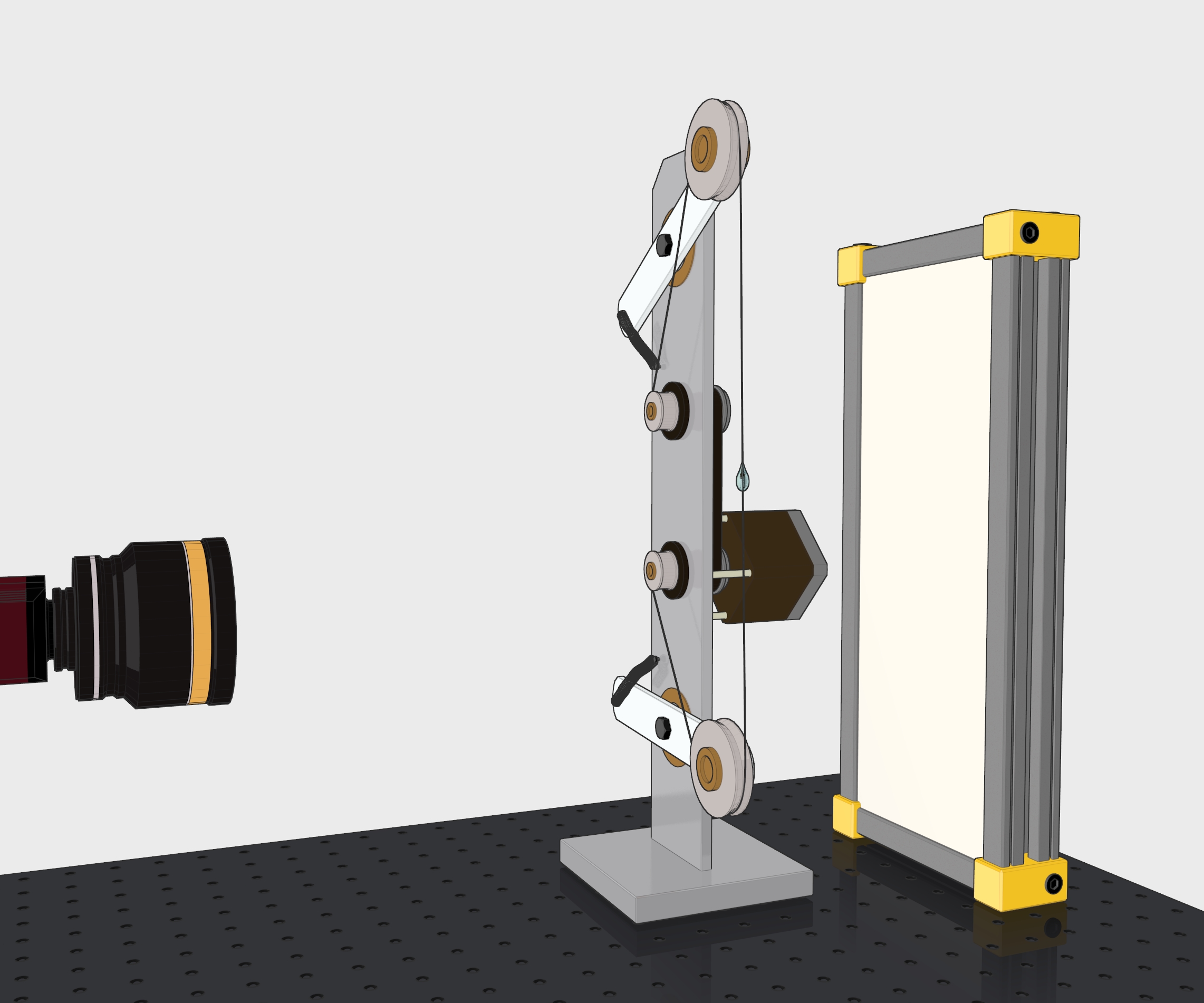}\\
(b) \includegraphics[width=0.45\textwidth]{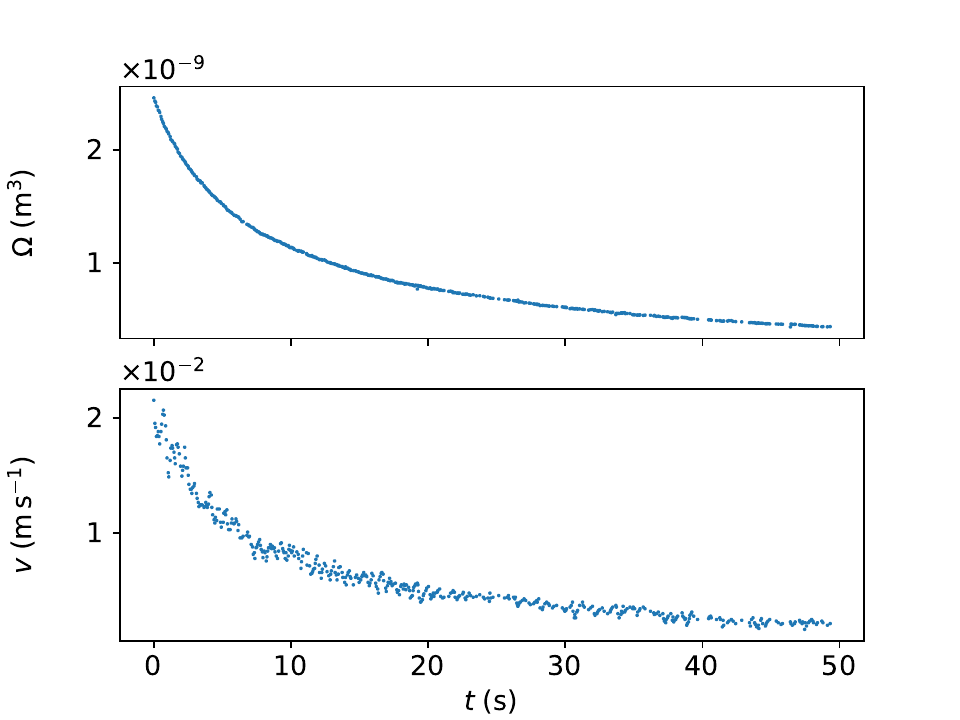}
\caption{(a) Sketch of the experimental setup where a camera records the droplet placed on a moving thread. Back illumination allows contrasted images. (b) Typical measurements of a droplet sliding on a single vertical fiber ($d=140 \, {\rm \mu m}$) : both volume $\Omega$ and speed $v$ are decreasing rapidly over time}.
\label{Figure_speed_volume}
\end{center}
\end{figure}

Experiments are performed with the help of an original setup sketched in Figure \ref{Figure_speed_volume}(a). The equipment considers a motorized fiber with an upward vertical motion in order to compensate the natural downward droplet motion. The experimental setup consists of two coils that are motorized and synchronized. Their purpose is to unwind the fiber from the lower coil to the upper one. Additionally, two external pulleys, which are held in place by springs, are used to maintain a tension in the fiber. The fiber itself is enlighted from the back and the motion of the droplets is recorded from the front thanks to a CCD camera (Charge-Coupled Device). The latter is connected to a computer which tracks the droplet in real time. A feedback loop adjusts the rotational speed of the spool motor to the droplet speed in order to keep the droplet at the center of the images. The speed $v$ of the droplet is therefore measured. A sufficient length of wire (approximately 3m) is wrapped around the lower coil. This allows for multiple experiments to be conducted before needing to rewind the wire in the opposite direction while cleaning the fiber with a tissue imbibed with isopropanol.

Moreover, the contour of the droplet is detected on each image. Thanks to the Pappus's centroid theorem \cite{Pappus}, the volume $\Omega$ of the droplet is estimated by integration, (i) assuming that the droplet keeps an axis symmetry around the fibers and (ii) by taking into account the volume of the portion of the fibers wrapped by the droplet. This method for measuring the volume is very good for single fibers, while becomes less reliable when droplet sizes becomes of the order of the bundle size. The shape characteristics of the droplet are also identified on the pictures : height $L$ and width $\ell$. Those lengths are sketched in Figure \ref{Figure_droplets}(a). Typical data of synchronous measurement of speed $v$ and volume $\Omega$ are shown in Figure \ref{Figure_speed_volume}(b). One observes a fast decrease of both quantities over time.

The motion of the droplet along a single fiber or the bundle of fibers is driven by gravity and surface tension as revealed by the following  non-dimensionnal numbers. We can estimate the Bond number ${\rm Bo} =\rho g \Omega/2\pi\gamma d_e$, which compares gravitational and capillary effects by taking the typical values of $\Omega$ in our experiments. We obtain ${\rm Bo}$ values around $0.75$ for moving droplets, meaning that gravity drive the system, as well as capillary effects. The capillary number ${\rm Ca} = v \eta /\gamma $ compares surface tension effects to viscous ones. We have ${\rm Ca}$ values between $0.05$ and $0.25$ revealing that surface tension may overcome viscosity. The Weber number ${\rm We}= \rho v^2 d_e/8\gamma$ is typically around $10^{-6}$ such that inertia can be neglected in the present study. This assumption has been considered in \cite{gilet2010droplets} where similar fiber sizes and drop volumes were considered.

\section{Experimental results}

In the following, we present results with three different viscosities and for $n=1$ to $n=4$ using the same color code for all figures. From low to high viscosities, data are colored in green, blue and red. From dark colors to light ones, the number of fibers is increased from $n=1$ to $n=4$. For each set of parameters, data are averaged over 3 to 5 experiments. 

\subsection{Droplet shape}

\begin{figure}[h]
\begin{center}
(a)\includegraphics[width=0.45\textwidth]{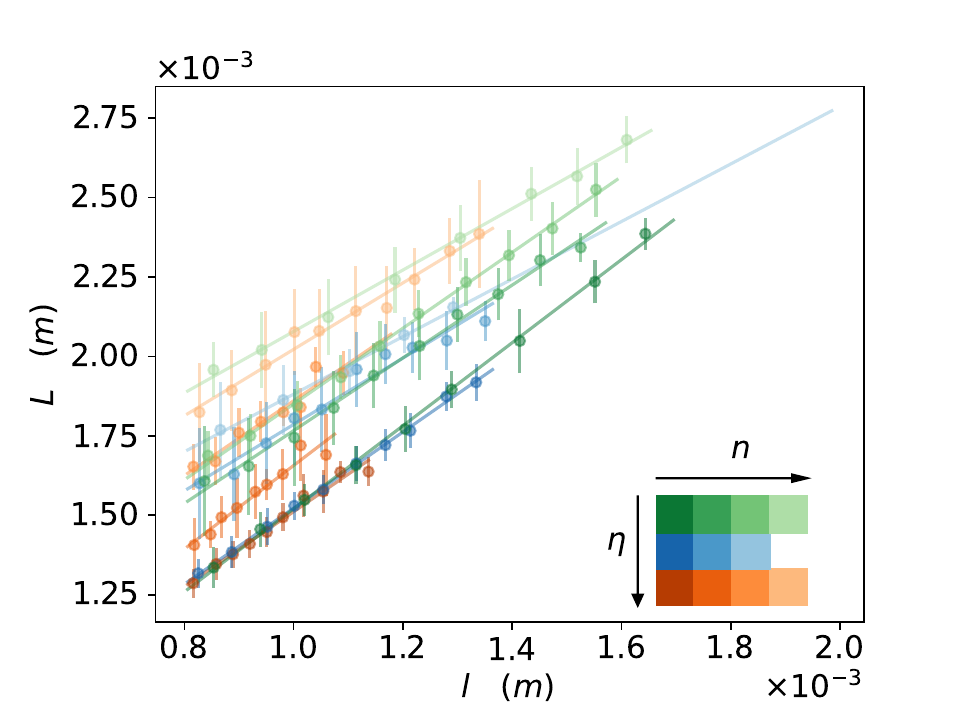}
(b)\includegraphics[width=0.45\textwidth]{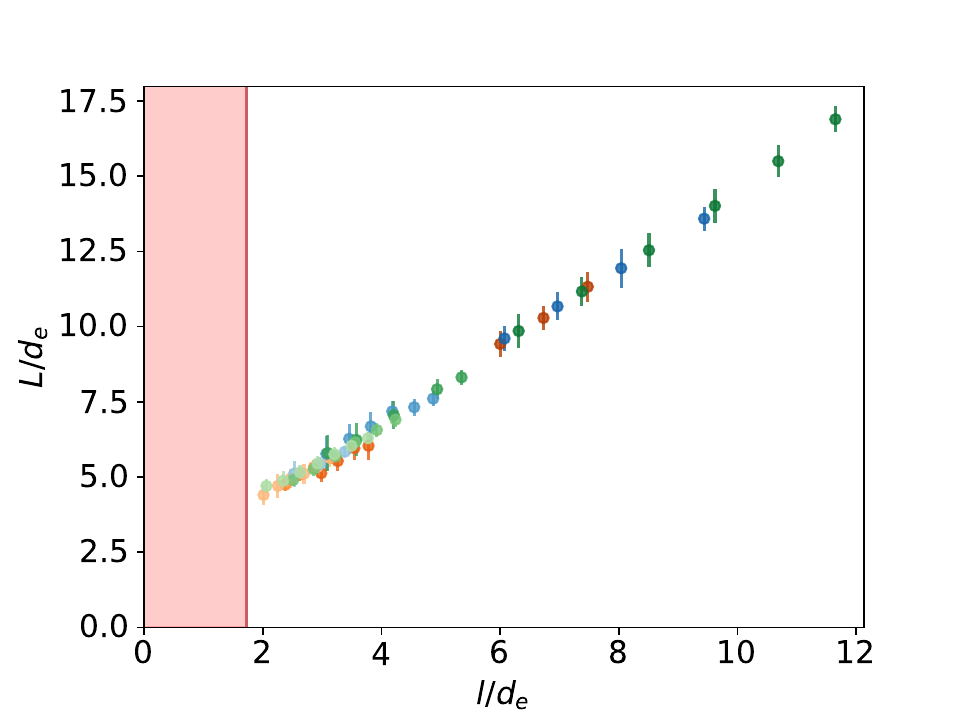}
\caption{(a) Height of the droplet $L$ as a function of the width $\ell$ of the droplet. All data points are colored as a function of viscosity $\eta$ and fiber number $n$. See the color legend given as an inset. (b) Lengths are rescaled by the effective diameter $d_e$. The region colored in red corresponds to the $\Omega_c < 1$ criterion such that a droplet there is no longer in a barrel configuration. }
\label{Figure_L_l}
\end{center}
\end{figure}

Figure \ref{Figure_L_l}(a) shows the width $\ell$ and height $L$ of droplets over all experiments in a single plot. Color codes correspond to various $n=\{1,2,3,4 \}$ values and various viscosities $\eta=\{10^{-2}, 2 \times 10^{-2}, 5\times 10^{-2} \} \, {\rm Pa \, s}$. Since the droplet volume $\Omega \approx \pi \ell^2 L / 6 - \pi d_e^2 L /4$ is decreasing during the experiments, geometrical characteristics $\ell$ and $L$ are evolving and are plotted in this figure. A linear behavior is found for each set of data points. The slope is close to unity meaning that droplet characteristic lengths are evolving in a similar way. There is however an offset which depends roughly on the fiber number $n$. The droplets seem more elongated when grooves are present. The spreading of a droplet is indeed enhanced on grooved substrates, as studied recently in \cite{VanHulle2023}.

By normalizing $\ell$ and $L$ by the effective thread diameter $d_e$, all data points collapse around a single linear behavior in Figure \ref{Figure_L_l}(b). This finding implies that $d_e$ is a pertinent parameter for characterizing both the shape and size of droplets, regardless of viscosity or fiber count. Furthermore, the collapse of the data highlights the fact that the number of grooves does not affect the geometry of the droplets.  It should be remarked that droplets on fibers are often considered in the scientific literature as nearly spherical objects \cite{carroll1986equilibrium,lorenceau2004drops,gilet2010droplets} with an aspect ratio close to unity. We however evidence herein aspect ratios in the range $L/\ell \approx 1.4 - 1.8$ in dynamical cases.

The red colored region in Figure \ref{Figure_L_l}(b) at small $\ell/d_e$ values is where droplets should leave their barrel shape to a clam shape. Indeed, it is possible to define the reduced volume, denoted by $\Omega_c$, which represents the ratio of the droplet's volume $\Omega$ to the volume of the fiber $\pi d_e^2 L /4$ enclosed by the droplet. The criterion for keeping a barrel shape droplet is given by $\Omega_c>1$ \cite{carroll1986equilibrium,van2021capillary}. Taking the characteristic lengths into account, this barrel shape criterion becomes
\begin{equation}
\frac{\ell}{d_e} > \sqrt{3}
\end{equation}
where all our data are seen. The lower limit $\ell = \sqrt{3} d_e$ is denoted by a red vertical line in Figure \ref{Figure_L_l}(b).

\subsection{Film thickness}

\begin{figure}[h]
\begin{center}
(a) \includegraphics[width=0.45\textwidth]{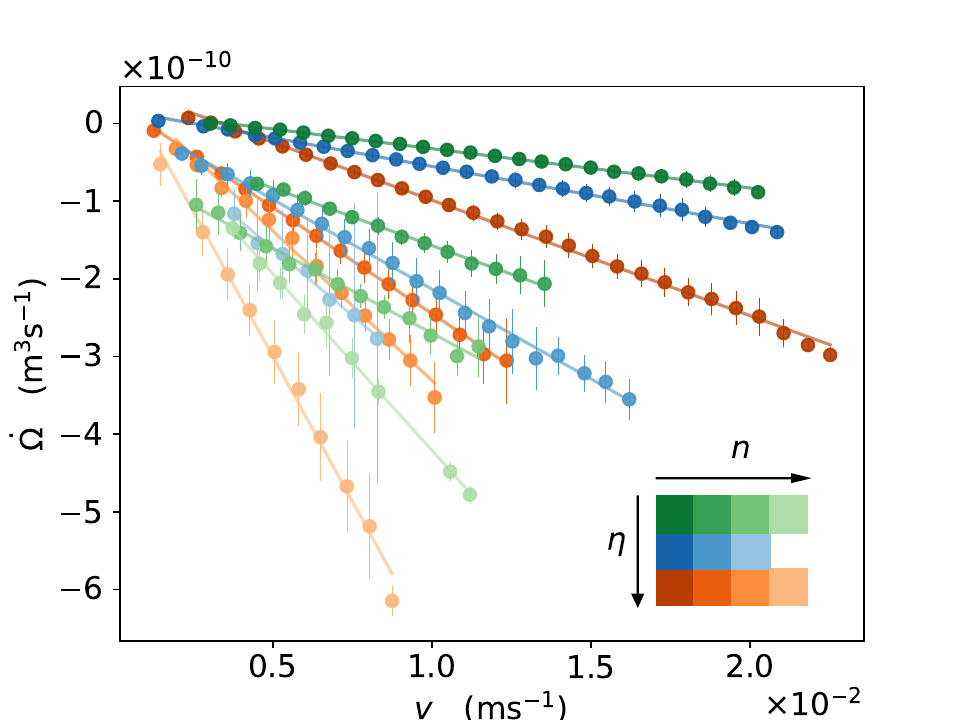}
(b) \includegraphics[width=0.45\textwidth]{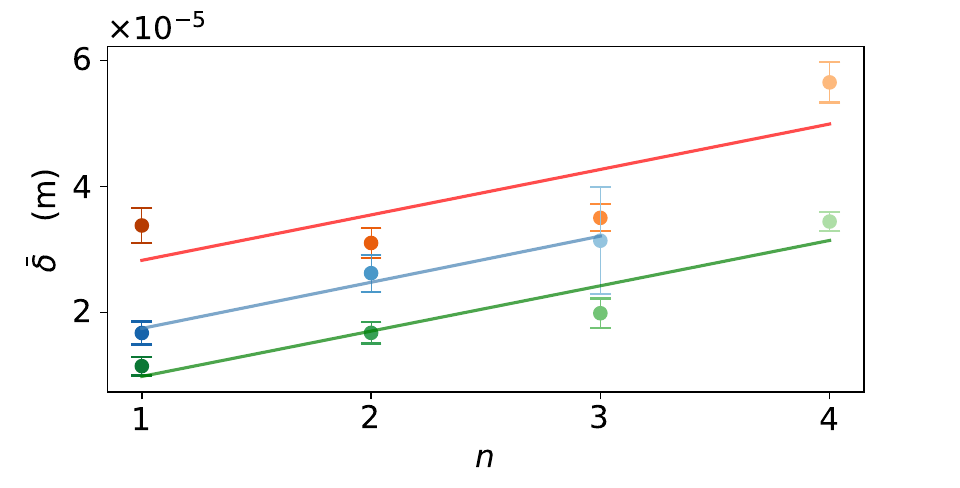}
(c) \includegraphics[width=0.45\textwidth]{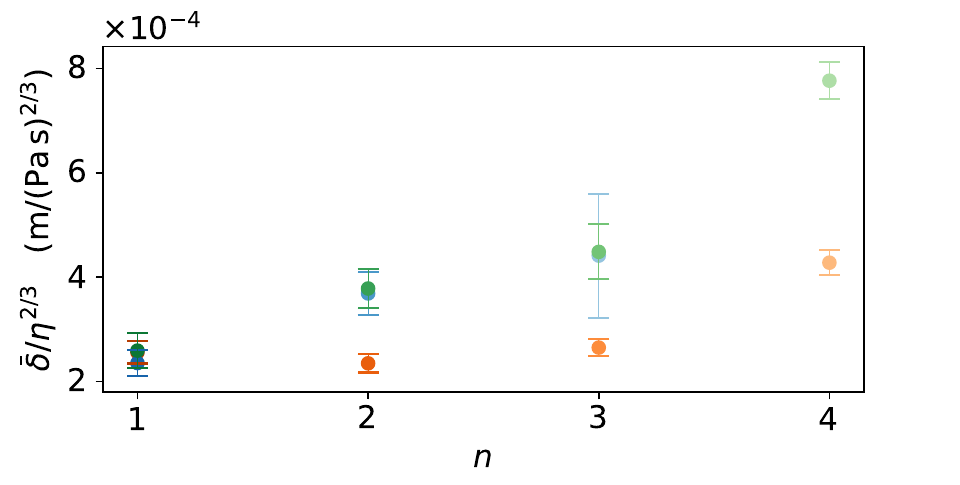}
\caption{(a) Volume loss rate $\dot \Omega$ as a function of droplet speed $v$. Lines are linear fits used to extract the slopes for determining $\bar \delta$ from Eq.(\ref{eq_volume}). (b) Average Liquid film thickness $\bar \delta$ as a function of $n$. Lines are fitted on the data to emphasize $\eta$ and $n$ dependencies, but should be considered as guides for the eye. (c) Average film thickness normalized by $\eta^{2/3}$ emphasizing that the effect of viscosity is captured by the Landau-Levich model only for $n=1$.}
\label{Figure_OmegaDot}
\end{center}
\end{figure}

The volume $\Omega$ is decreasing as the droplet slides along the fiber, leaving a thin film behind. 
Liquid volume conservation states that 
\begin{equation}
\dot \Omega = - \pi d_e \delta v.
\label{eq_volume}
\end{equation}
where $\delta \ll d_e$ is the thickness of the liquid film left behind the droplet, as observed and sketched in Figure \ref{Figure_droplets}. Whatever the $\eta$ and $n$ values, a nearly linear behavior is obtained in Figure \ref{Figure_OmegaDot}(a) when $\dot \Omega$ is plotted as a function of speed $v$. All results are shown following the same color code as the inset of Figure \ref{Figure_L_l}. Different slopes are observed, depending on viscosity $\eta$ and number $n$ of fibers. 

From the slopes measured in Figure \ref{Figure_OmegaDot}(a), the relationship Eq.(\ref{eq_volume}) allows us to estimate the average film thickness $\bar \delta$ left behind droplets. Results are plotted in Figure \ref{Figure_OmegaDot}(b) as a function of $n$. Typical values for $\bar \delta$ ranges from 10 to 50 $\rm \mu m$. The film thickness is seen to be highly dependent on liquid viscosity $\eta$. It should also be remarked that the presence of grooves ($n>1$) has also an important effect on the film thickness. Indeed, a clear increase of $\bar \delta$ is observed for all viscosities in Figure \ref{Figure_OmegaDot}(b). In fact, liquid amounts are accumulated into the grooves. The above behaviors are emphasized in Figure \ref{Figure_OmegaDot}(b) by colored lines which are guides for the eye. 

Moreover, the Landau-Levich model \cite{landau1988dragging} of liquid coating is predicting a film thickness $\delta \propto {\rm Ca}^{2/3}$, and more precisely 
\begin{equation}
    \delta = 0.67 d_e {\left(\eta v \over \gamma \right)}^{2/3}
\end{equation}
as proposed in \cite{quere1999fluid}. The volume loss vs speed should therefore scale as $\dot \Omega \propto -v^{5/3}$. However, the range of speed values (or Ca values) is less than a decade such that a nearly linear behaviour is observed in Figure \ref{Figure_OmegaDot}(a) instead. Higher speeds cannot be reached with this experimental setup. Nevertheless, the liquid coating is highly sensitive to viscosity and one expects that the film thickness increases as $\bar \delta \propto \eta^{2/3}$. For a single fiber ($n=1$), the three measured thicknesses are in agreement with the Landau-Levich model since $\bar \delta/\eta^{2/3}$ gives a unique value of $2.5\times 10^{-4} \,\rm{m/(Pa\,s)^{2/3}}$ within error bars. The cases with grooves ($n>1$) are different since the penetration of the liquid into grooves relies heavily on viscosity, and here the Landau-Levich model ceases to provide accurate predictions.

\subsection{Droplet speed}

\begin{figure}
\centering
(a) \includegraphics[width=0.45 \textwidth]{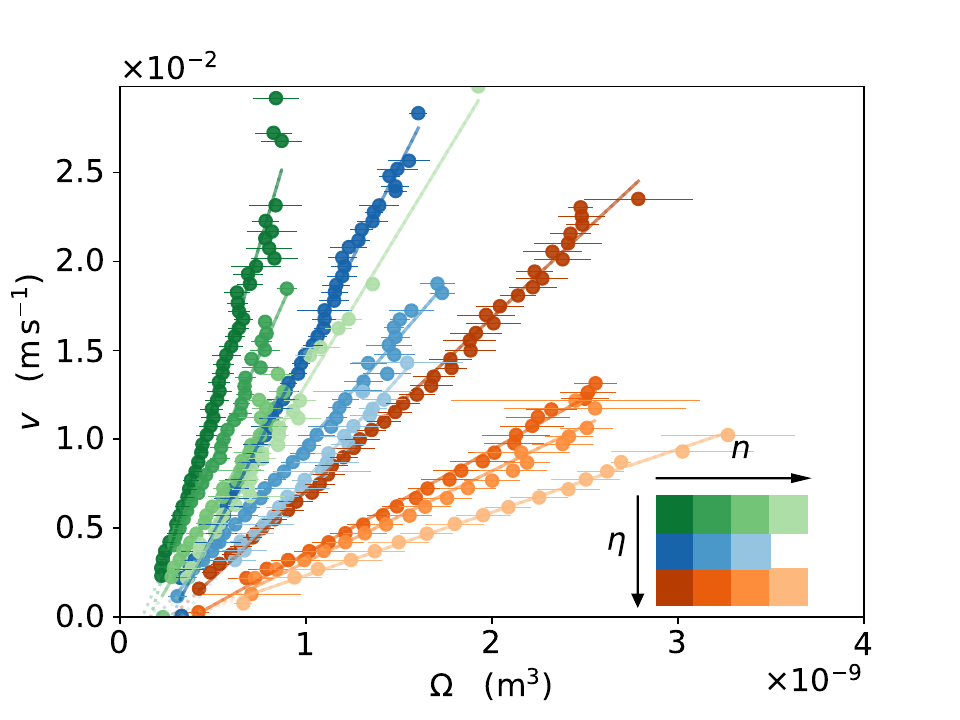}
(b) \includegraphics[width=0.45\textwidth]{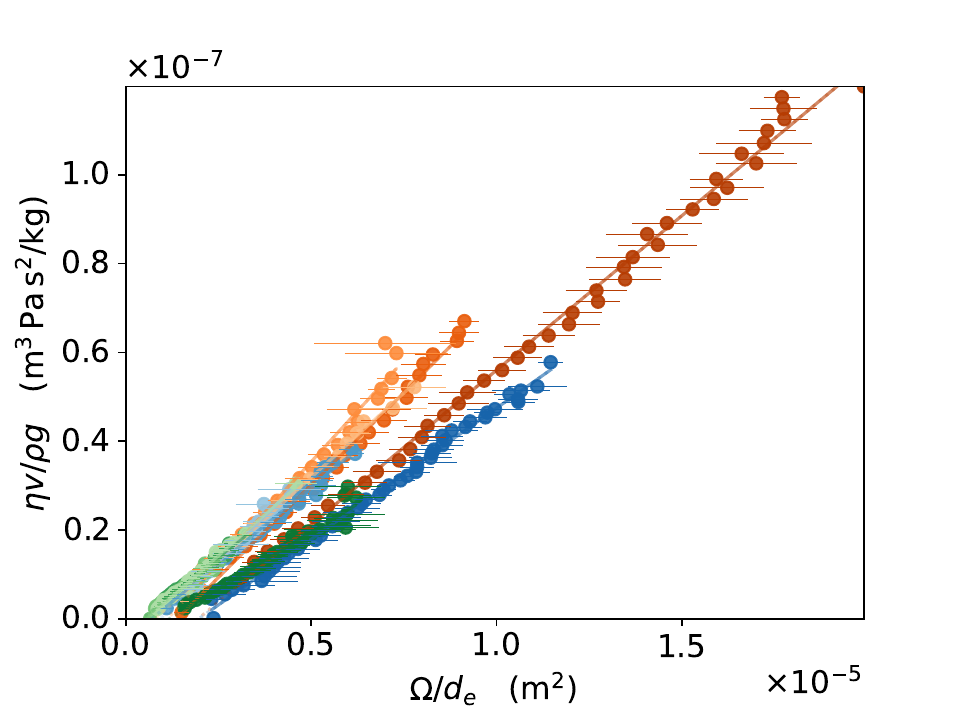}
(c) \includegraphics[width=0.45\textwidth]{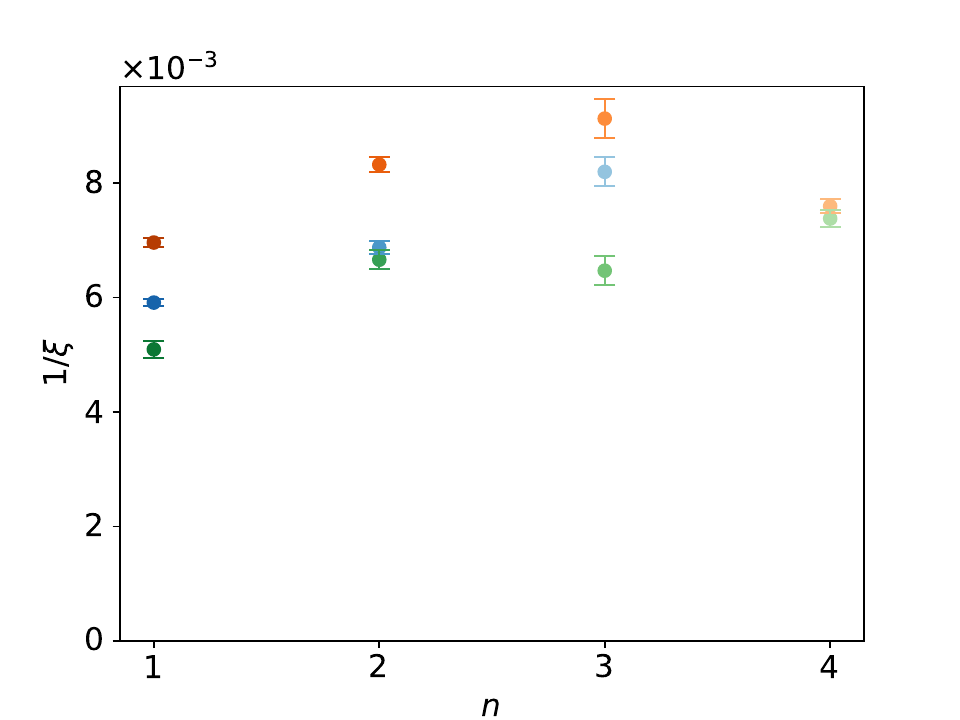}
\caption{(a) Droplet speed as a function of $\Omega$ emphasizing the gravity driven mechanism. Each color is associated to viscosity $\eta$ and fiber number $n$. Error bars are given. Data are fitted by a linear model of Eq.(\ref{eq_speed}). (b) Rescaled speed $v$ and volume $\Omega$ according to Eq.(\ref{eq_speed}). (c) From the slopes extracted in the top graph, the inverse dissipation factor $1/\xi$ is estimated. It is plotted as a function of $n$ with the same color code. One observes that the presence of grooves favors the droplet motion, i.e. $1/\xi$ increases with $n$. }
\label{fig_Omegaspeed}
\end{figure}

In Figure \ref{fig_Omegaspeed}(a), the speed $v$ of the droplet is plotted as a function of the measured volume $\Omega$. One may notice that the speed increases with the volume, a result which is consistent since the weight of the droplet is the driving force of the problem. A linear behavior is observed between both physical quantities. Increasing the effective fiber diameter $d_e$ reduces the speed, and more precisely the slope of the linear behavior. Also, different oil viscosities have been used in order to evidence the role of dissipation mechanisms. Increasing the viscosity has an effect on the speed which is reduced accordingly. On top of that observation, one remarks that all linear trends are converging towards a particular volume at zero speed. It means that below that particular volume $\Omega_0$, tiny droplets are static on the vertical fiber. This particular volume seems more dependent on fiber diameter than on viscosity, as we will see below. 

\section{Discussion}

\subsection{Droplet dynamics}

In order to capture the linear behaviors of Figure \ref{fig_Omegaspeed}(a), i.e. the droplet dynamics, we propose a model based on the sum of three forces acting on the droplet : gravity, dissipation and capillary. The driving force is the weight of the droplet and is given by $F_g=\rho g \Omega$, where $g$ is the gravitational acceleration. 

Dissipation should play an important role and following \cite{gilet2010droplets}, we consider a classical drag force $F_{d} = -\xi \eta d_e v$, where $\xi$ is a dissipation factor. Assuming that the friction force $F_d$ is mainly due to velocity gradients within the droplet, and following \cite{lorenceau2004drops,gilet2010droplets}, one has 
\begin{equation}
F_d = - \pi d_e \int_{a_f}^{L/2} c \, {\eta v \over z} \, {L \over \ell} \, dz + 
\pi d_e \int_{L/2}^{a_r} c \, {\eta v \over z} \, {L \over \ell} \, dz
\end{equation}
where velocity gradients $v/z$ are integrated from the shortest distance from the contact line at either the front $a_f$ or the rear $a_r$ of the droplet. We cut the droplet in equal parts $L/2$ although the lack of front/rear symmetry, because the sources of dissipation are mainly located at the front and at the rear of the droplet. The dimensionless constant $c$ is a geometrical factor close to unity studied in \cite{gilet2010droplets}. After integration over the front of the droplet and the rear of the droplet, the main contributions to $\xi$ are due to the front and rear characteristics, and more precisely 
\begin{equation}
\xi = c \pi {L \over \ell} \ln \left( {L^2 \over 4 a_f a_r}\right)
\label{eq_xi}
\end{equation}
In front of the droplet, the fiber is dry such that $a_f$ is estimated to $a_f \approx 10^{-9}$ m from deGennes \cite{deGennes1985wetting} while at the rear of the droplet this distance is limited by the film thickness $a_r=\delta$. Injecting geometrical characteristics $L/\ell$ as measured herein and $\bar \delta$ around 10 $\mu$m, one obtains $\xi \approx 115$.

The capillary force is acting on both sides of the droplet. In the front, i.e. at the bottom of the droplet, the contact line separates a dry fiber and the droplet, such that the contribution to the capillary force is given by $\pi \gamma d_e$. Indeed, silicone oil is totally spreading on the fiber such that contact angle can be considered as zero. Behind the droplet, capillary force comes from the liquid film. One has a contribution $-\pi (d_e+\delta) \gamma$. Summing both contributions give a capillary force $F_c =-\pi \delta \gamma$.

Therefore, taking into account for the sum of the forces acting on the droplet, the equation of motion becomes
\begin{equation}
\rho \Omega \dot{v} = \rho g \Omega - \pi \delta \gamma -  \xi \eta d_e v .
\label{eq_motion}
\end{equation}
Neglecting inertia, as suggested at the end of Section II, the speed of the droplet $v$ is given by 
\begin{equation}
v = \dfrac{\rho g \Omega}{\xi \eta d_e} - \dfrac{ \pi\gamma \delta}{\xi \eta d_e} .
\label{eq_speed}
\end{equation}
One finds the linear dependency of the speed with the droplet's volume with a single slope. Figure \ref{fig_Omegaspeed}(b) plots normalized quantities $\eta v / \rho g$ as a function of $\Omega / d_e$ to evidence such linear behavior. Different slopes are however observed, indicating that the dissipation factor depends on $\eta$ and $n$. This will be discussed in the next subsection. 

Setting the speed to zero in Eq.(\ref{eq_speed}), one can find the volume offset $\Omega_0$ being
\begin{equation}
\Omega_0 = \dfrac{\pi \delta \gamma}{\rho g}.
\label{eq_offset}
\end{equation}
Injecting the values of surface tension and density in the last relationship, one finds a typical volume offset around 0.1 $\mu \ell$, in agreement with our measurements. Please note that the above condition can be interpreted in terms of Bond numbers. Indeed, Eq.(\ref{eq_offset}) can be rewritten into ${\rm Bo_c} = \bar \delta/2d_e \approx 0.05$ meaning that all droplets characterized by a larger Bond number are moving until they loose volume and reach this critical low ${\rm Bo_c}$ value.

\subsection{Dissipation factor}

By fitting the slopes of Figure \ref{fig_Omegaspeed}(b) with Eq.(\ref{eq_speed}), the inverse dissipation factor $1/\xi$ can be extracted from all experiments. The results are shown in Figure \ref{fig_Omegaspeed}(c). One observes that the dissipation factor (and therefore droplet speed) is dependent on viscosity $\eta$ and fiber number $n$. The values of $1/\xi$ are close to the one estimated in the previous subsection, i.e. $1/\xi \approx 1/115 = 8.7 \times 10^{-3}$. Surprisingly, the inverse dissipation factor seems more important for intermediate $n$ values such that higher speeds are observed for fiber bundles. The effect is already present for $n=2$ and $n=3$. There, a significant speed increase of about 20\% is obtained from the single fiber case. This effect is counter intuitive.

In Figure \ref{fig_Omegaspeed}(c), one observes that $\xi$ decreases with $\eta$ and is quite sensistive to the substructure, i.e. depends on $n$. Indeed, Eq.(\ref{eq_xi}) links $\xi$ to $\delta$. Since high viscosities imply thicker films behind the droplet, one expects less dissipation. A second observation is that substructure has a significant effect on dissipation since $\xi$ is lower for $n=2$ and $n=3$ than for $n=1$. Of course, the previous argument on film thickness could be also used. Grooves allow for faster droplet motion due to thicker films behind. It is hard to separate effects but it should be noted that the groove effect is significant with 20\% extra speeds for some configurations. 

\begin{figure}[h]
\begin{center}
(a)
\includegraphics[width=0.25\textwidth]{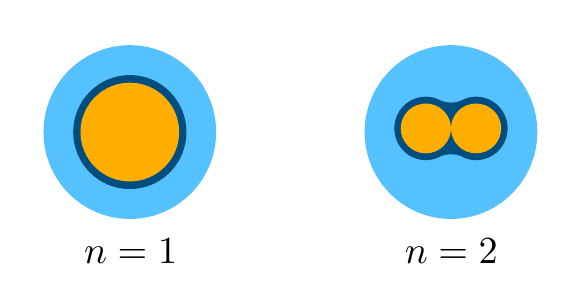}\\
(b)
\includegraphics[width=0.45\textwidth]{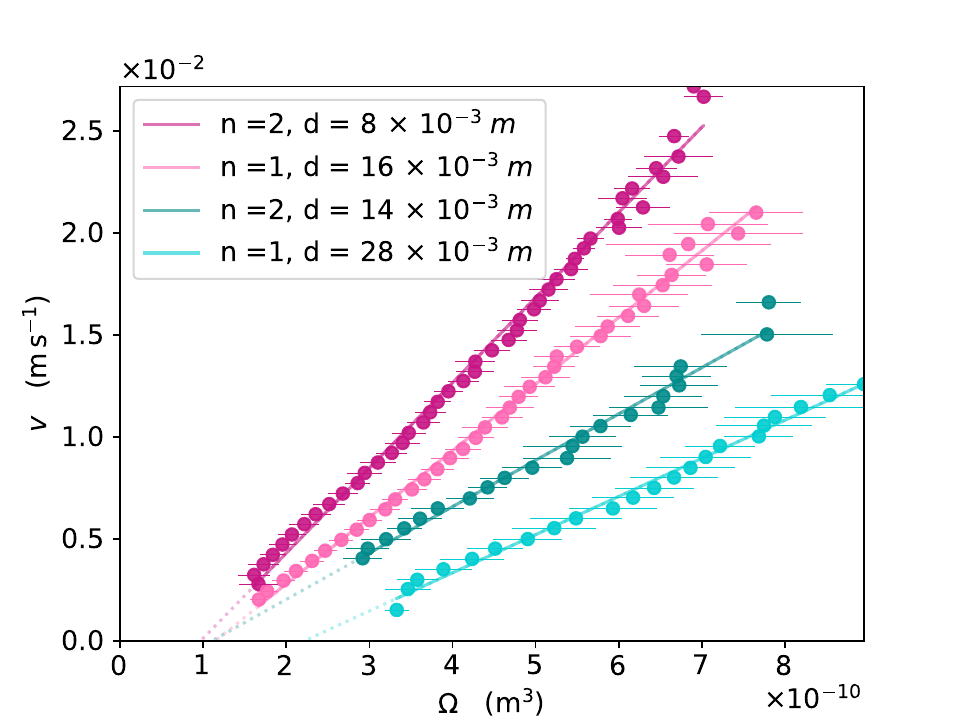}
\caption{(a) Sketch of the additional experiment where a single fiber of diameter $d$ is compared with $n=2$ fibers of diameter $d/2$. The equivalent diameter $d_e$ is therefore conserved but grooves are present in the second case. (b) Droplet speed as a function of droplet volume for two situations illustrated in (a). Extra speed is clearly obtained for $n=2$. }
\label{diam_eq}
\end{center}
\end{figure}

To assess the effect of grooves, we performed additional experiments. In figure \ref{diam_eq}, droplet dynamics are shown for comparing the case of single fiber of diameter $d$ with 2 adjacent fibers of diameter $d/2$, thus fixing the equivalent diameter $d_e=d$. Two different diameters $d$ are tested. In both cases, the droplet speed $v$ is much higher when $n=2$ than for $n=1$. These additional experiments confirm extra speed for 2 joined fibers equivalent to a single one.

A last remark is that the dynamics emphasized hereabove leads to a more complex phenomenon on longer times. Indeed, any droplet is losing volume and finally stops when reaching $\Omega_0$. After a while, the coating left above the immobile mother droplet forms smaller daughter droplets due to the classical Rayleigh-Plateau instability. Those daughter droplets are on a prewetted fiber such that they start moving as soon as they grow. Finally they reach the mother droplet. Fed by daughter droplets, the mother droplet starts again its motion till reaching again $\Omega_0$. A start and stop motion of the primary droplet develops. Future works will focus on this phenomenon. 


\section{Conclusion}

In this paper, we built an experimental setup able to measure simultaneously droplet speed $v$ and volume $\Omega$ during their motion on various threads. The dynamics of the droplet motion was captured and models were proposed. Such measurements allow for estimating the dissipation factor $\xi$ which is dependent on viscosity because of the film thickness left behind. 

On top of that, we evidenced that the droplet motion is deeply affected by the substructure of the vertical thread. Indeed, grooves are collecting more liquid such that the film thickness left behind the droplet depends on the groove number. As a consequence, dissipation is modified such that extra speeds (up to 20\%) are observed with threads possessing grooves than smooth cases. 

The findings of this study have practical implications for water harvesting applications, as droplet velocity along fibers has a direct impact on the efficiency of water capture. Vertical fiber-based structures, known as harps, have shown great potential in water harvesting \cite{shi2020harps} and efficiently drain water after a certain onset time \cite{Jiang2019}. The onset time refers to the time required for the first captured droplets to begin draining and is a primordial characteristic in weather changing conditions. The use of fiber bundles, as described in this article, has the potential to reduce the onset time of these structures. 

Another application concerns characterization methods. Indeed, our setup is able to extract relevant information about the dissipation inside droplets. The system can therefore be used as a type of rheometer. Such a device will be developed in the future and could be tested with other liquids like non-Newtonian and complex liquids. 

\vskip 2mm
\section*{Acknowledgments} 

This work has been partially supported by the FNRS-WISD project number X.3047.17. The authors thanks M.Mélard for his precious help developing the experimental setup.

\vskip 0.5cm

%
\end{document}